\documentclass[a4paper,11pt]{article}
\usepackage{pos}

\newcommand{\bec}{\begin{center}}
\newcommand{\eec}{\end{center}}
\newcommand{\beq}{\begin{equation}}
\newcommand{\eeq}{\end{equation}}
\newcommand{\bea}{\begin{eqnarray}}
\newcommand{\eea}{\end{eqnarray}}

\newcommand{\hf}{\frac{1}{2}}

\newcommand{\cN}{{\cal N}}

\title{Complex Langevin simulations for $PT$-symmetric models}

\author*[a]{Arpith Kumar}
\author[a]{Anosh Joseph}

\affiliation[a]{Department of Physical Sciences, \\Indian Institute of Science Education and Research (IISER) Mohali, \\Knowledge City, Sector 81, SAS Nagar, Punjab 140306, India}

\emailAdd{anoshjoseph@iisermohali.ac.in}
\emailAdd{arpithk.iiserm@gmail.com}

\abstract{Self-interacting scalar quantum field theories possessing $PT$-symmetry are physically admissible since their energy spectrum is real and bounded below. However, models with $PT$-invariant potentials can have complex actions in general and a non-perturbative study of such systems using methods based on traditional Monte Carlo is hindered due to numerical sign problem. In this work we employ complex Langevin based on stochastic quantization to study two-dimensional scalar field theories, including the ones exhibiting $PT$-symmetry. We also study the simplest supersymmetric version of these systems and address the question on dynamical supersymmetry breaking.
}

\FullConference{%
 The 38th International Symposium on Lattice Field Theory, LATTICE2021
  26th-30th July, 2021
  Zoom/Gather@Massachusetts Institute of Technology
}

\begin{document}
\maketitle

\section{Introduction and motivation}

A systematic investigation of non-perturbative regimes of quantum field theories can be performed with the help of lattice regularization. Traditional numerical methods based on Monte Carlo can in general provide the physics of such systems when the Euclidean action is real. However, in theories with complex actions, path integral Monte Carlo encounters the notorious {\it sign-problem} that makes the simulation algorithms unreliable. In the literature there exists only a handful of methods that can effectively tackle the sign-problem. We study quantum field theory models in two dimensions with the help of one of these methods - the complex Langevin method \cite{Klauder:1983nn, Parisi:1983mgm}.

Quantum field theories such as QCD with quark chemical potential, QCD with a topological $\theta$-term, Chern-Simons gauge theories, and chiral gauge theories can suffer from the sign-problem. The list also includes an interesting class of non-Hermitian and self-interacting quantum field theories that exhibit $PT$-invariance. Although formulated using complex actions, these theories can possess real and bounded below energy spectra. In Refs. \cite{Bender:1997ps, Bender:1998gh} Bender and Milton considered a new class of $PT$-invariant (Euclidean) quantum field theories with interactions of the form $\lambda( i \phi )^{(2 + \delta)}$. These theories are physically admissible, that is, they possess a real and bounded below energy spectra. But for these interactions parity in itself is manifestly broken. Our goal is to perform a non-perturbative analysis of these theories with the help of complex Langevin method. 

Incorporation of supersymmetry (SUSY) into our understanding of particle physics has drawn considerable interest among physicists ever since it was first proposed. Our experiences show that SUSY can only be a fundamental symmetry of nature if it is manifested as a spontaneously broken symmetry at low energy scales \cite{Witten:1981nf, Witten:1982df}. This leads to the requirement that in order to study SUSY breaking mechanisms we need non-perturbative tools. Over the past few decades, a lot of effort has been put into formulating lattice regularized supersymmetric models, and thus providing access to various intriguing non-perturbative phenomena such as dynamical SUSY breaking. 

In this proceedings we present our preliminary investigations, with the help of complex Langevin method, on the study of lattice regularized version of a minimal supersymmetric model, namely $\mathcal{N} = 1$ Wess-Zumino model in two-dimensions. We discuss the case when the superpotential is a double-well potential. We also discuss our ongoing simulations of the $PT$-symmetric superpotentials. Before moving on to supersymmetric models in Sec. \ref{sec:susy_models}, as a warm up, in Sec. \ref{sec:Two-dimensional_scalar_field_theories} we discuss two-dimensional scalar field theories with $\phi^4$ and $PT$-symmetric potentials.

\section{Two-dimensional scalar field theories}
\label{sec:Two-dimensional_scalar_field_theories}

Consider the Lagrangian of a two-dimensional Euclidean scalar field theory 
\beq
\mathcal{L}_E = \hf \partial_\mu \phi  \partial_\mu \phi +\hf m^2 \phi^2 + W(\phi),
\eeq
where $\phi$ is a dimensionless scalar, $m$ is the mass parameter, and $W(\phi)$ is the interaction potential. The Euclidean action is $S_E = \int d^2 x~\mathcal{L}_E$. 

To simulate the model using the complex Langevin method we first discretize the model on a two-dimensional toroidal lattice. The temporal and spatial extents, $\beta_t$ and $\beta_x$, respectively, can be expressed as $\beta_t = \beta_x = La$ with $L$ denoting the number of lattice sites in each direction and $a$ denoting the lattice spacing. We have $\int d^2 x \longrightarrow a^2 \sum_x$. The periodicity of the lattice enables us to write
\beq
\left(\partial_\mu \phi \right)^2 = - \phi \partial_\mu^2 \phi = -\frac{1}{a^2} \left[ \phi_{x}\phi_{x+\mu} + \phi_{x}\phi_{x-\mu} -2{\phi_{x}}^2\right],
\eeq
where $\phi_{x\pm \mu}$ represents the field at the neighboring site in $\pm \mu$-th direction. 

Using the complex Langevin method we can study these models for various interaction potentials including the $PT$-invariant potentials. Complex Langevin update for field configurations at a lattice site $x$, for Langevin time $\theta$, with step-size $\epsilon$ is given by
\beq
\phi_{x, \theta + \epsilon} = \phi_{x, \theta} + \epsilon v_{x, \theta} + \eta_{x, \theta} \sqrt{\epsilon}, 
\eeq
where the drift term is obtained as $v_{x, \theta} = - { \partial S_E }/{ \partial \phi_{x, \theta} }$ and $\eta_{x, \theta}$ is a real Gaussian noise.

\subsection{Model with $\phi^4$ potential}

Consider the potential $W(\phi) = \lambda \phi^4$. Classically, the model is invariant under the discrete $\mathbb{Z}_2$ symmetry, that is, $\phi \to -\phi$. However, in quantum theory, this symmetry may be broken dynamically. The expectation value of the scalar field, $\langle \phi \rangle$ can be regarded as an order parameter. If $\langle \phi \rangle = 0$ the theory is in a symmetric phase, otherwise it is in a symmetry broken phase. There exist comprehensive studies of the $\phi^4$ theory on the lattice \cite{De:2005ny, Schaich:2009jk, Wozar:2011gu, Sakai:2018xkx}. We will utilize this model as a testbed for our Langevin analysis. We employ a lattice parameterization with dimensionless lattice parameters $m_0^2 = m^2 a^2$ and $\lambda_0 = \lambda a^2$, and in addition, we introduce a new set of parameters $\kappa$ and $\tilde{\lambda}$ \cite{De:2005ny}, 
\beq
m_0^2 \to \frac{1-2\tilde{\lambda}}{\kappa} -4, ~~ \lambda_0 \to 6\frac{\tilde{\lambda}}{\kappa^2}, ~~ {\rm and} ~~ \phi \to \sqrt{2\kappa} \Phi.
\eeq
The above parameterization leads to the lattice action 
\beq
S = - 2 \kappa \sum_x \sum_\mu \Phi_x \Phi_{x + \mu} + \sum_x \Phi_x^2 + \tilde{\lambda} \sum_x \left( \Phi_x^2 - 1 \right)^2.
\eeq

\begin{figure}
	\centering
	\includegraphics[width=.329\textwidth]{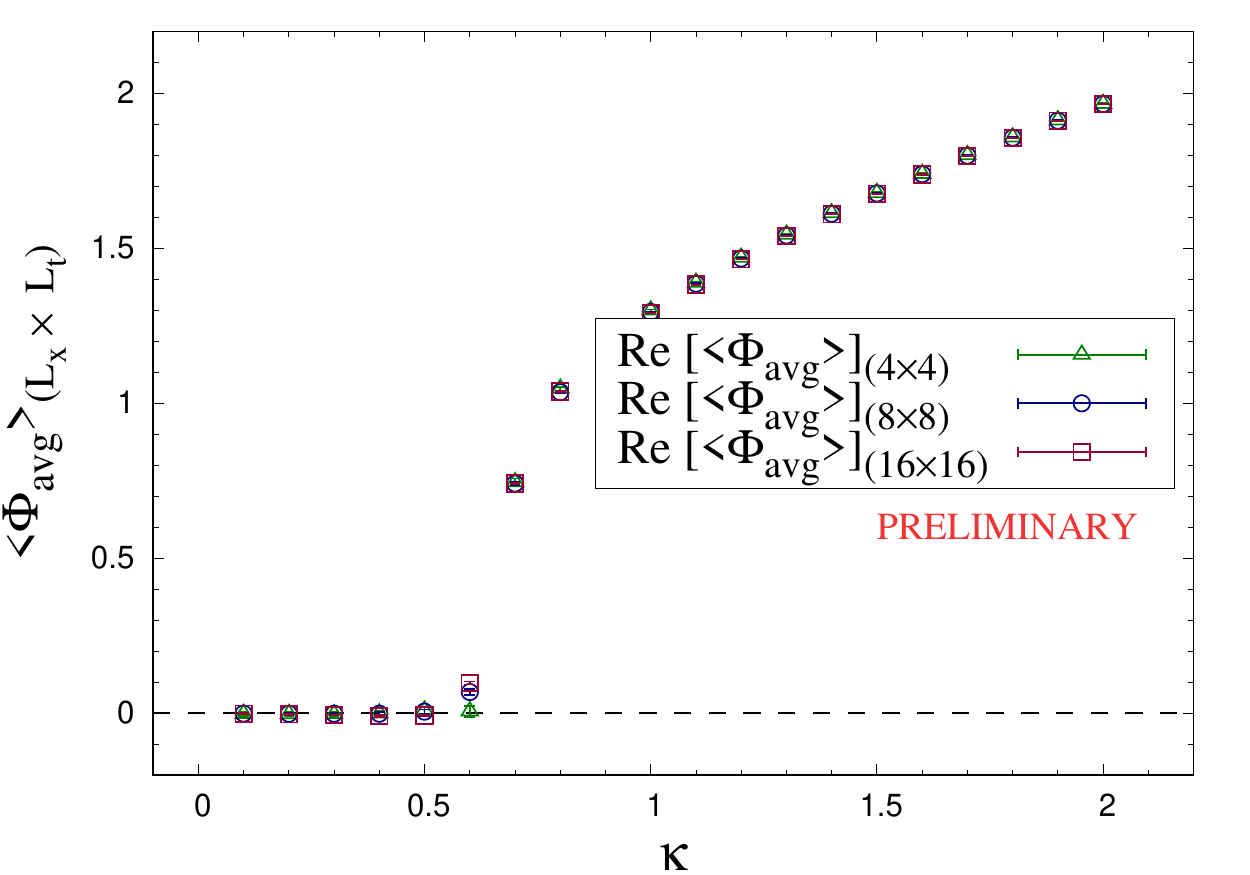}
	\includegraphics[width=.329\textwidth]{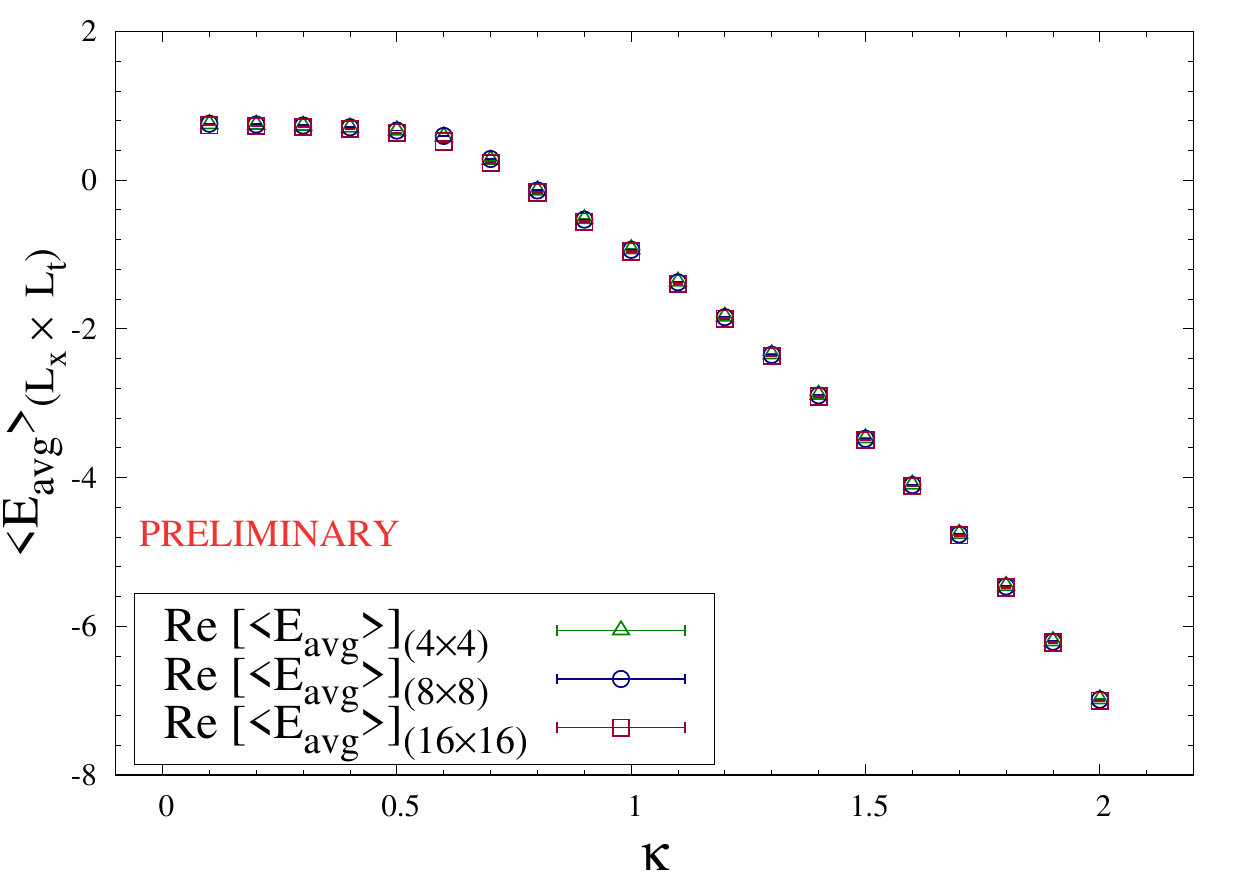}
	\includegraphics[width=.329\textwidth]{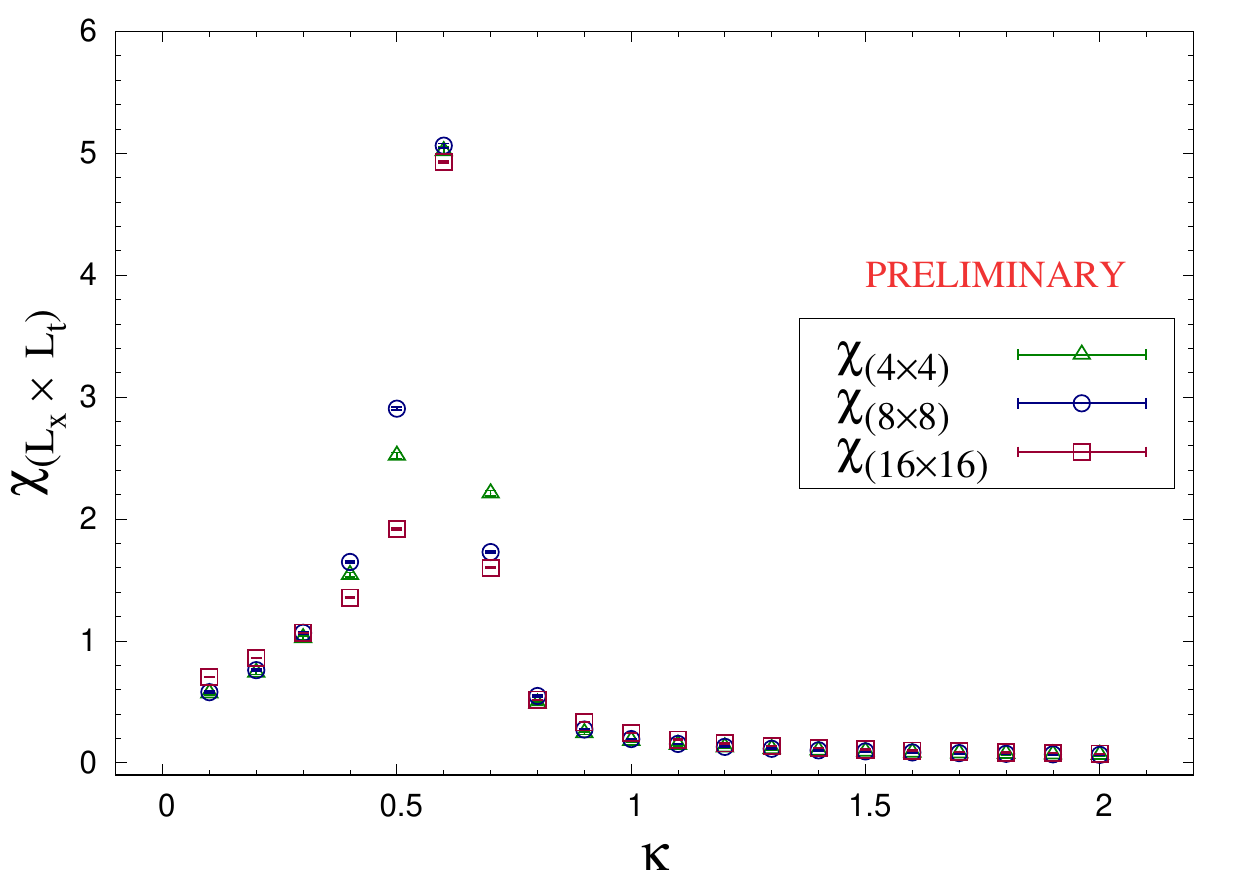}
	\caption{Model with $\phi^4$ potential. Expectation values of the order parameter $\Phi$ (left panel), energy $E$ (center panel), and susceptibility $\chi$ (right panel) against $\kappa$ for different lattice extents and fixed $\tilde{\lambda} = 0.5$.}
	\label{fig:phi4}
\end{figure}

In our simulations of the model we monitor the following observables as $\kappa$ is varied: the average of the field $\Phi$ as an order parameter, energy $E$, and susceptibility $\chi$. The simulation results are shown in Fig. \ref{fig:phi4} for different lattice extents and fixed $\tilde{\lambda} = 0.5$. The results indicate that the model possess a phase transition around $\kappa = 0.6$ and $ \langle  \Phi_{\rm avg} \rangle  \neq 0$ for $\kappa \ge 0.6$ implying $\mathbb{Z}_2$ broken phase.

\subsection{Model with $PT$-invariant potential}

Next we move onto $PT$-invariant scalar field theory with the potential $W(\phi) = - \lambda (i\phi)^{(2 + \delta)}$, where the coupling $\lambda$ has $m^2$ dimension and $\delta$ is a real parameter. It is fascinating to note that these models possess a real and bounded below spectra for $\delta > 0$ with a non-zero mass parameter. The positivity of the spectrum can be understood from a theoretical point of view. 

As an example, we consider the theory for $\delta = 1$. The Lagrangian is
\beq
\mathcal{L}_E = \hf \left(\partial_\mu \phi\right)^2 +\hf m^2 \phi^2 + i \lambda \phi^3.
\eeq
For a conventional real $\lambda \phi^3$ theory, in the weak-coupling expansion, the Green's functions can be expressed as a formal power series in $\lambda^2$. This power series, although real, does not alternate in sign, and hence, is not Borel summable. The non-summability of perturbation series reflects the fact that the spectrum is not bounded below. Upon replacing the coupling $\lambda$ $\to$ $i\lambda$, the theory becomes $PT$-symmetric. The power series remains real and also it alternates sign. As a consequence the perturbation series becomes summable suggesting that the underlying theory possesses real positive spectrum \cite{Bender:1997ps, Bender:1998gh, Milton:2003av}.

The action for such $PT$-symmetric theories is complex in general. Path integral Monte Carlo requires the action to be real and hence a non-perturbative lattice study of these theories is hindered due to a sign problem or \textit{complex phase problem}. We use complex Langevin method to overcome this difficulty. For the $\delta = 1$ model, the lattice action can be expressed as
\beq
S = -\sum_x \sum_\mu \phi_x \phi_{x + \mu} + \left( 2 + \frac{m_0^2}{2} \right) \sum_x \phi_x^2 + i \lambda_0 \sum_x \phi_x^3,
\eeq
where $m_0$ and $\lambda_0$ are dimensionless mass and coupling parameters, respectively.

\begin{figure}
	\centering
	\includegraphics[width=.45\textwidth]{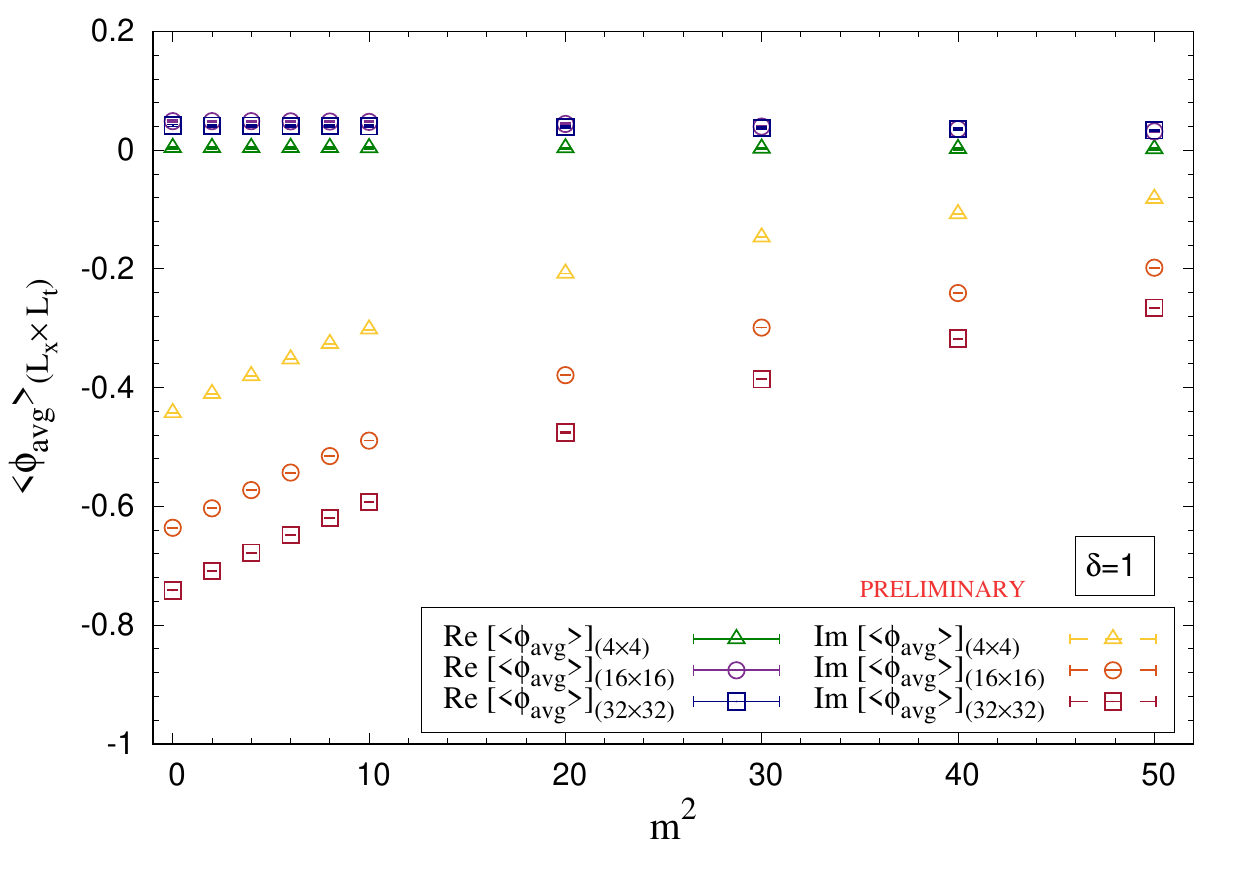}
	\includegraphics[width=.45\textwidth]{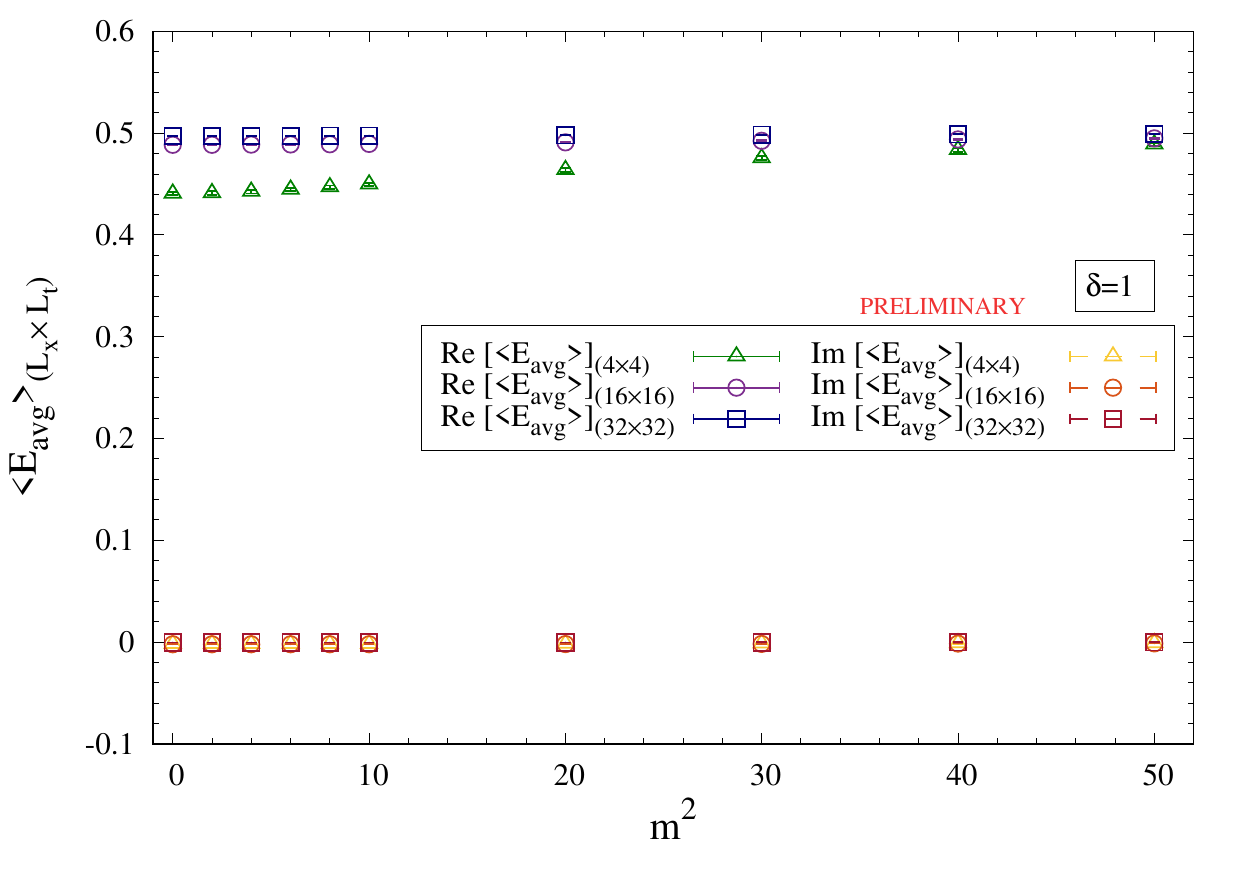}

	\includegraphics[width=.45\textwidth]{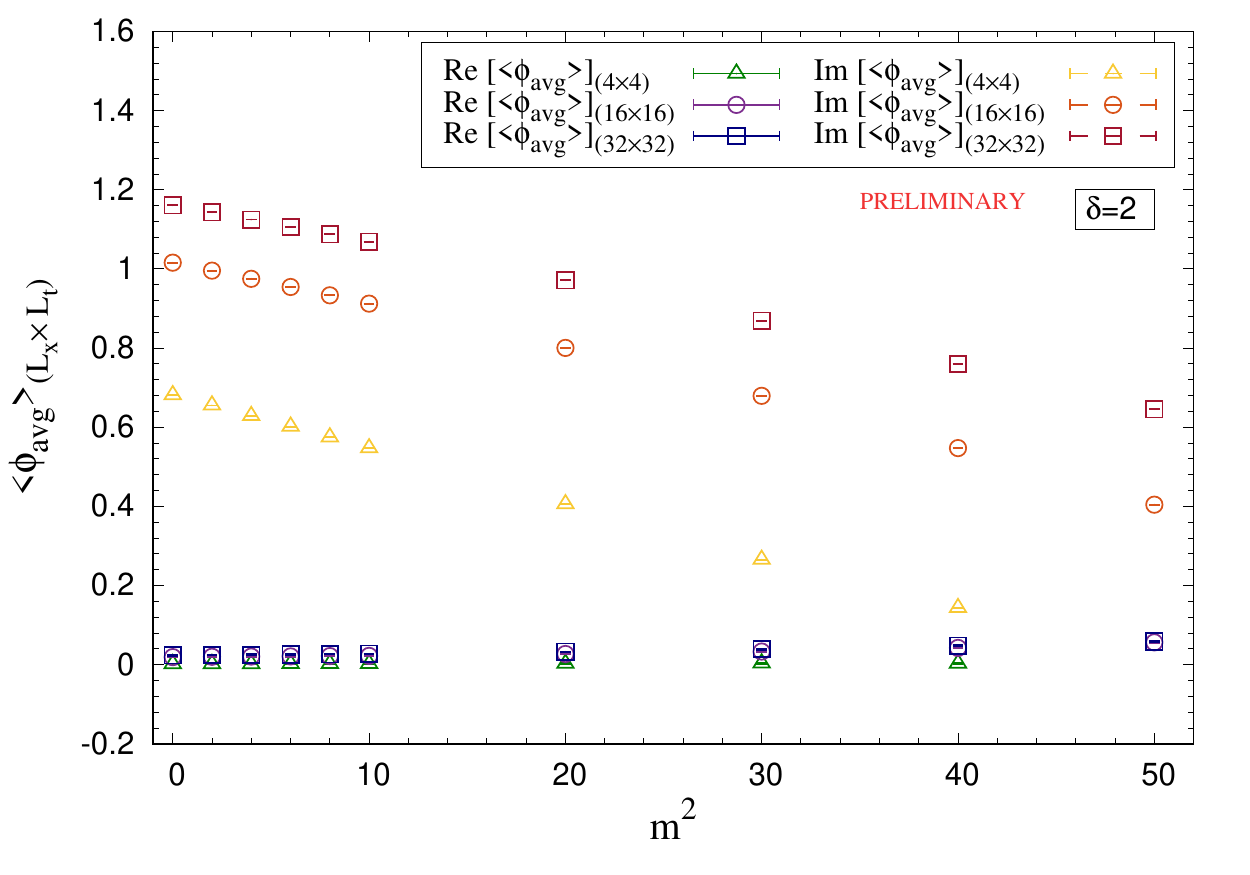}
	\includegraphics[width=.45\textwidth]{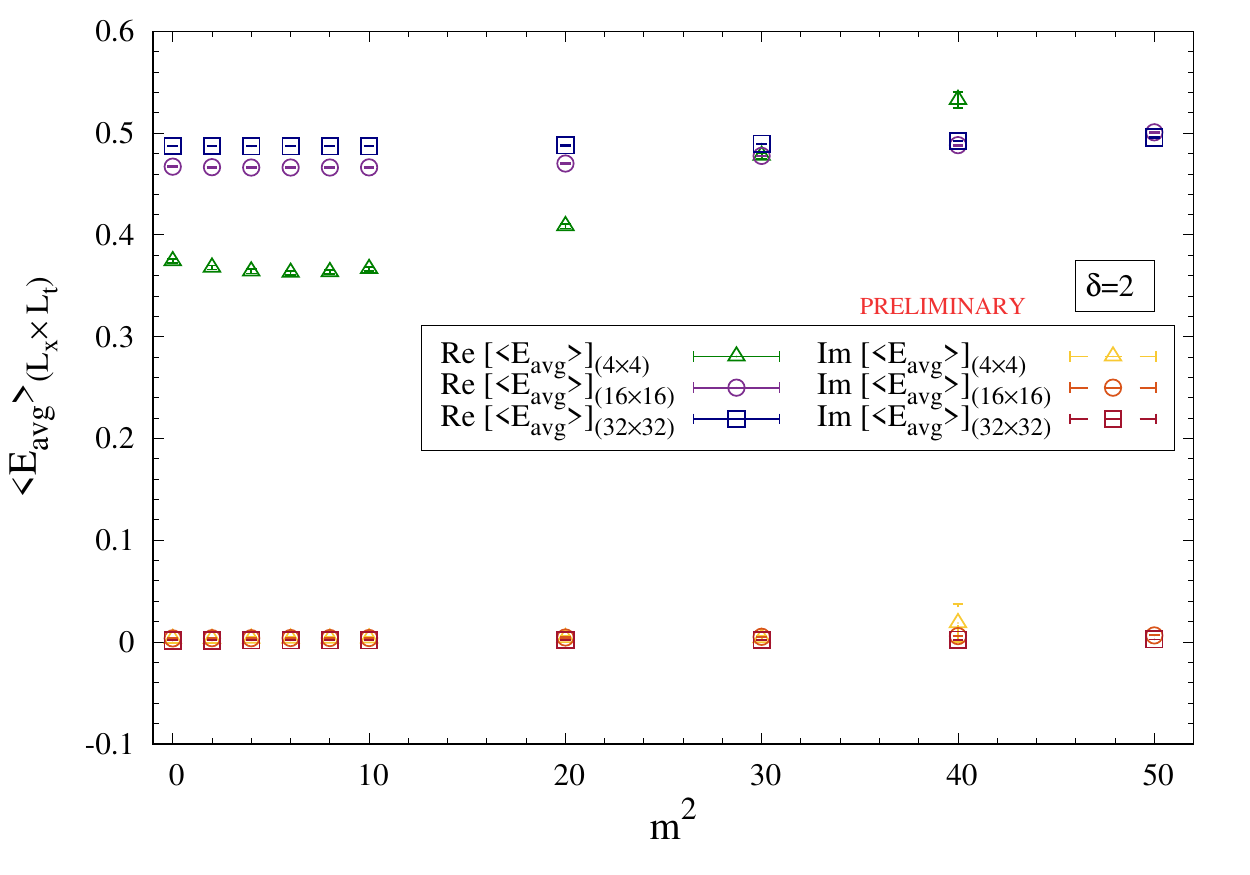}

	\caption{Bosonic $PT$-symmetric model with $\delta = 1$ (top) and $\delta = 2$ (bottom) potential. The expectation values of the real and imaginary parts of the order parameter $\phi$ (left panel) and the energy $E$ (right panel) against physical mass parameter $m^2$ for different lattice extents and fixed physical coupling $\lambda = 10.0$.}
	\label{fig:pt1-2}
\end{figure}

In Fig. \ref{fig:pt1-2} we show our simulation results for the bosonic $PT$-symmetric theory with $\delta = 1$ (top) and $\delta = 2$ (bottom) potential. On the left panel, the expectation values of the real and imaginary parts of the average field $\phi$ (order parameter) against physical mass $m^2$ for different lattice extents and fixed physical coupling $\lambda = 10.0$ is shown. On the right panel we show the ground state energy $E$ against $m^2$ for different lattice extents and fixed physical coupling $\lambda = 10.0$. These preliminary results suggest $ \langle  \phi_{\rm avg} \rangle  \neq 0$, that is, parity is manifestly broken for $\delta=1,~2$. The expectation value of energy is real and positive, Re [$\langle {\rm E_{avg}} \rangle] > 0$ and Im $[\langle {\rm E_{avg}} \rangle] = 0$, indicating a real bounded below spectra for this class of interactions. Our simulation results are in accordance with the analytical predictions \cite{Bender:1998gh}.

\section{Two-dimensional $\mathcal{N} = 1$ Wess-Zumino model}
\label{sec:susy_models}

In this section we study a supersymmetric version of the model discussed in the previous section. (The zero- and one-dimensional cousins of this model were studied recently in Refs. \cite{Joseph:2020gdh, Joseph:2019sof}.) We add fermions to the Lagrangian and consider the simplest two-dimensional supersymmetric quantum field theory, the $\mathcal{N} = 1$ Wess-Zumino model. The theory involves a minimalistic set of fields, that is, a scalar field $\phi$ and a two-component Majorana spinor $\psi$. The on-shell model in Euclidean space-time has the action 
\beq
S_E = \int d^2 x ~\hf \left[ \left(\partial_\mu \phi\right)^2 + \bar{\psi} \mathcal{M} \psi + W^2\left( \phi \right) \right],
\eeq
where $\mathcal{M} = \gamma^\mu \partial_\mu + W'\left(\phi \right)$ is referred to as the fermion matrix and the potential $W(\phi)$ is actually the derivative of the superpotential. The action is invariant under a single supersymmetry given by the transformations
\beq
\delta \phi = \bar{\epsilon} \psi, ~~\delta \psi = \big[ \gamma^\mu \partial_\mu \phi- W\left(\phi \right)  \big] \epsilon, ~~ \delta \bar{\psi} = 0.
\eeq
The Majorana spinor satisfies the relation, $\bar{\psi} = \psi^T \mathcal{C}$, where $\mathcal{C}$ is the `charge conjugation' operator in Euclidean space. It is given as
\beq
C =	\begin{pmatrix}
	0 & -1 \\
	1 & 0 
\end{pmatrix}.
\eeq 
It is crucial to note that this theory does not have a $\mathcal{Q}$-exact formulation or a Nicolai map. The model is not obtained by dimensional reduction unlike its $\mathcal{N} = 2$ supersymmetric version. Another interesting property is that for periodic boundary conditions for fermions, dynamical breaking of SUSY is possible, that is, the vanishing of the Witten index, $\Delta = 0$ can happen. 

For a non-perturbative analysis of the model, we place the theory on a symmetric toroidal lattice discussed in the previous section. We consider a particular lattice formulation of the model introduced by Golterman and Petcher \cite{Golterman:1988ta}. After integrating out the fermions, the lattice representation of the Euclidean continuum action has the following bosonic and fermionic components
\beq
S = S_b + S_f, ~~S_b = \hf \left( - \phi_r \Box^2_{rr'} \phi_{r'} + W^2_r   \right) ,~~S_f = \ln \left[{\rm Pf}\mathcal{M}\right]  = - \hf {\rm tr} \left[ \ln \mathcal{M} \right],
\label{eqn:n1wz-Slat}
\eeq
where $r, r'$ are the lattice vectors, fermion matrix $\mathcal{M} \equiv \mathcal{M}^{\alpha \beta}_{r r'} = \gamma^{\mu}_{\alpha \beta} \mathcal{D}^{\mu}_{r r' } + \delta_{\alpha \beta} W'_{rr'}$, ${\rm Pf} \mathcal{M}$ is the Pfaffian of the fermion matrix. We use symmetric difference operators defined as follows
\bea
\mathcal{D}^\mu_{rr'} &=& \hf \left[\delta_{r + e_\mu , r' } - \delta_{r - e_\mu, r' } \right]  \\
\Box^n_{rr'} &=& \hf \sum_\mu \left[\delta_{r + ne_\mu, r' } + \delta_{r - ne_\mu, r' } - 2 \delta_{rr' } \right].
\eea  
Since the action Eq. \eqref{eqn:n1wz-Slat} can be complex in general, we apply the complex Langevin method to study the theory for various superpotentials. We use the Euler discretized Langevin equation for the $s$-th lattice vector at Langevin time $\theta$. The drift term is defined as
\beq
v_{s, \theta} = - \frac{\partial S}{\partial \phi_{s, \theta}} = \Box^2_{s r'} \phi_{r', \theta} - W_{r'} W'_{r' s} + \left( \frac{\partial \mathcal{M}}{\partial \phi_s} \right)^{\alpha \beta}_{rr'} \left( \mathcal{M}^{-1} \right)^{\beta \alpha}_{r'r}.
\eeq

In order to test the reliability of complex Langevin simulations we check the correctness criteria \cite{Nagata:2016vkn} based on the decay of the distribution $P(u)$ of the absolute value $u$ of the drift term. We have at a particular Langevin time $\theta$, the drift-term magnitude
$u_\theta = \sqrt{\left( {1}/{L^2} \right) \sum_s \left| v_{s,\theta} \right|^2}$. We can trust the simulations if the distribution $P(u)$ of $u$ falls off exponentially or faster.

\subsection{Double-well superpotential}

We begin with considering a quadratic interaction potential or a double-well superpotential \cite{Catterall:2003ae, Baumgartner:2011jw, Wozar:2011gu} of the form
\beq
W(\phi) = \lambda \phi^2 -\frac{m^2}{4\lambda}, ~~~ \lambda \neq 0. 
\eeq 
The theory has two classical vacua at $\phi = \pm m/2\lambda$. In the lattice theory, we consider dimensionless couplings $\lambda_0$ and $m_0$, related to their continuum counter parts through $\lambda_0 = \lambda a$ and $m_0 = ma$. The potential and its derivative take the following form
\beq
W_r = \lambda_0 \phi_r^2 - \frac{m_0^2}{4\lambda_0} -\hf \Box^1_{rr'} \phi_r,~~
W'_{r r'} \equiv \frac{\partial P_r}{\partial \phi_{r'}} = 2 \lambda_0 \phi_r \delta_{rr'} - \Box^1_{r r'},
\eeq
where $\Box^1_{r r'}$ is the Wilson mass operator, which vanishes in the continuum limit but eliminates fermion doubling problem at a finite lattice spacing. Due to the introduction of the Wilson term, the lattice action is no longer invariant under parity, implying that the two vacuum states are not equivalent. It is expected that field configurations would reside in the vicinity of one of the classical vacua.	In the large values of ${m_0}^2/\lambda_0$ the $Z_2$ symmetry is spontaneously broken (in infinite volume) and $\phi$ settles down to a definite ground state.

\begin{figure}
	\centering
	\includegraphics[width=.54\textwidth,origin=c,angle=0]{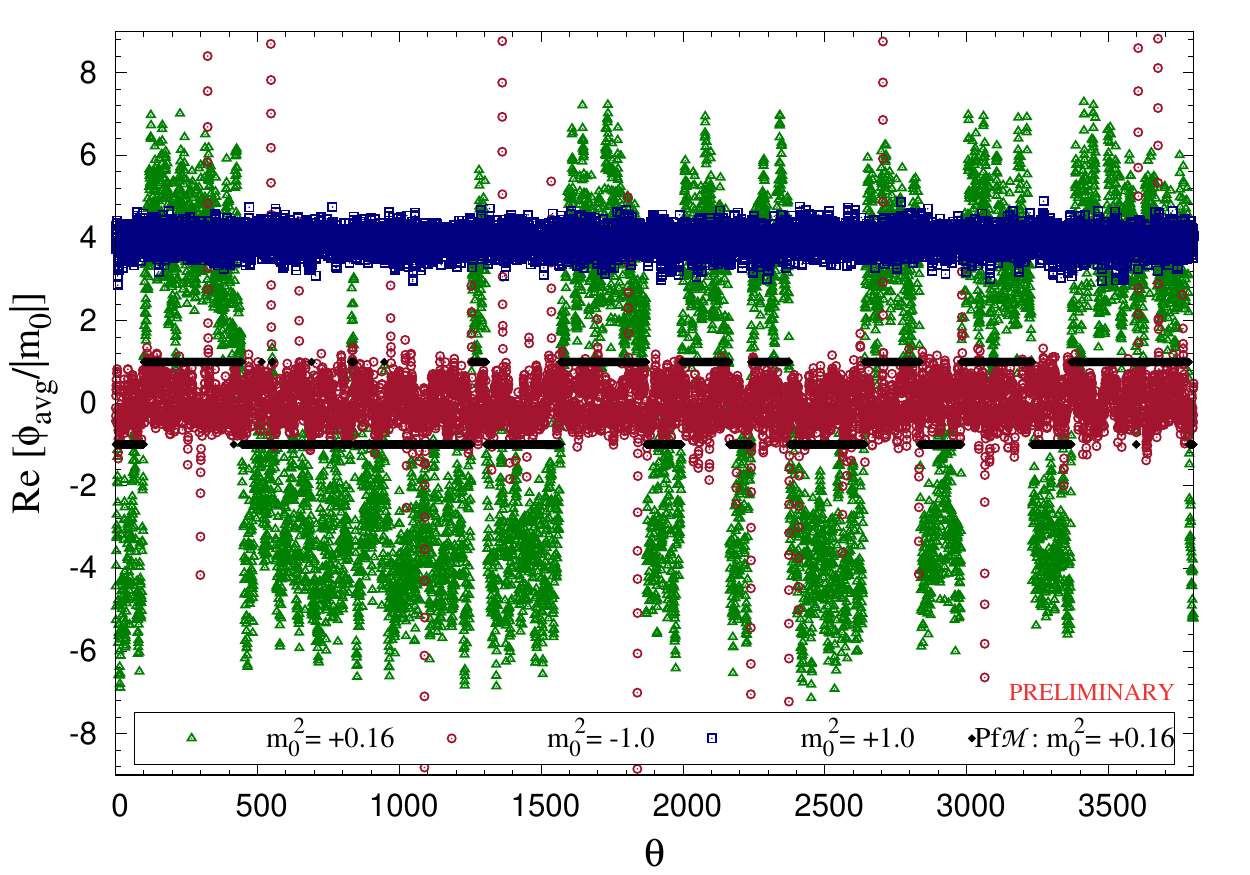}

	\includegraphics[width=.48\textwidth,origin=c,angle=0]{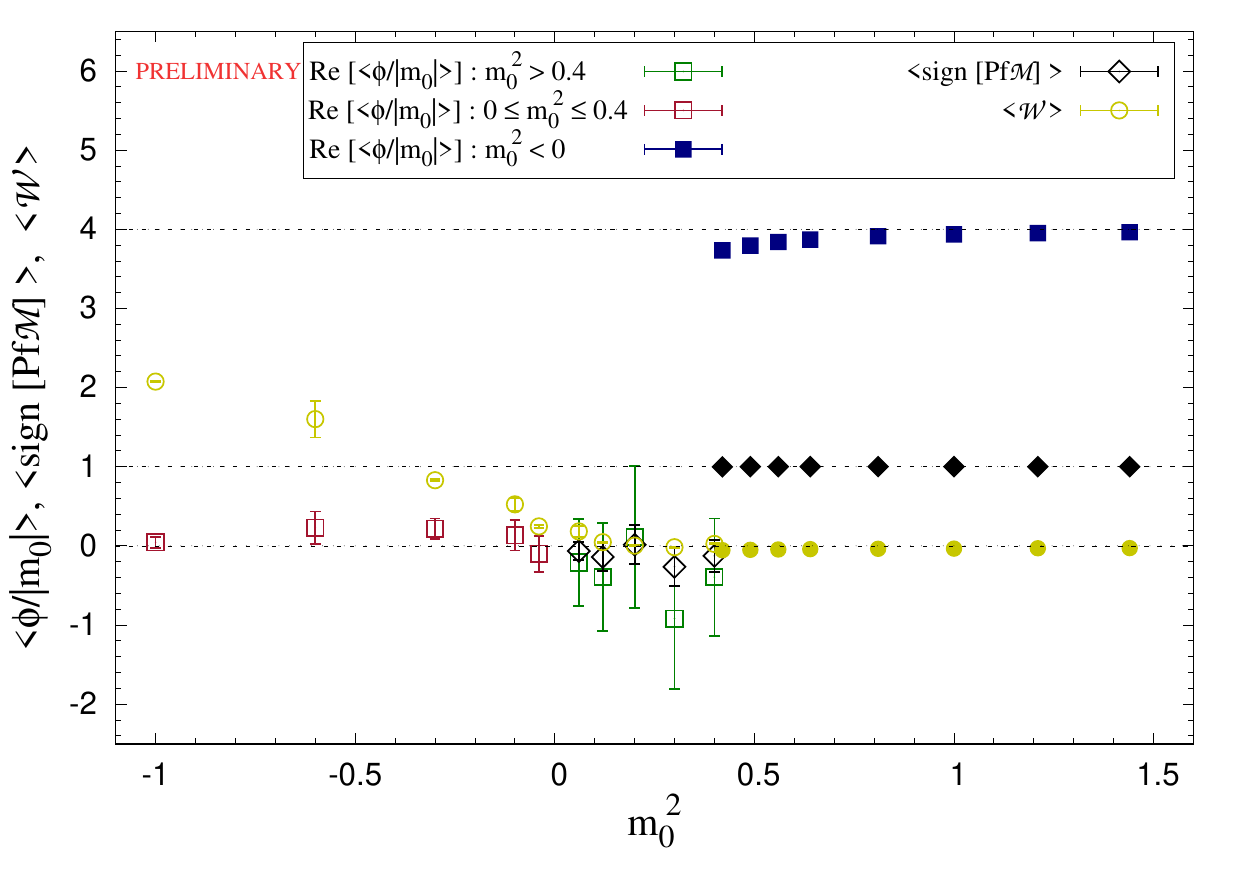}	
	\includegraphics[width=.48\textwidth,origin=c,angle=0]{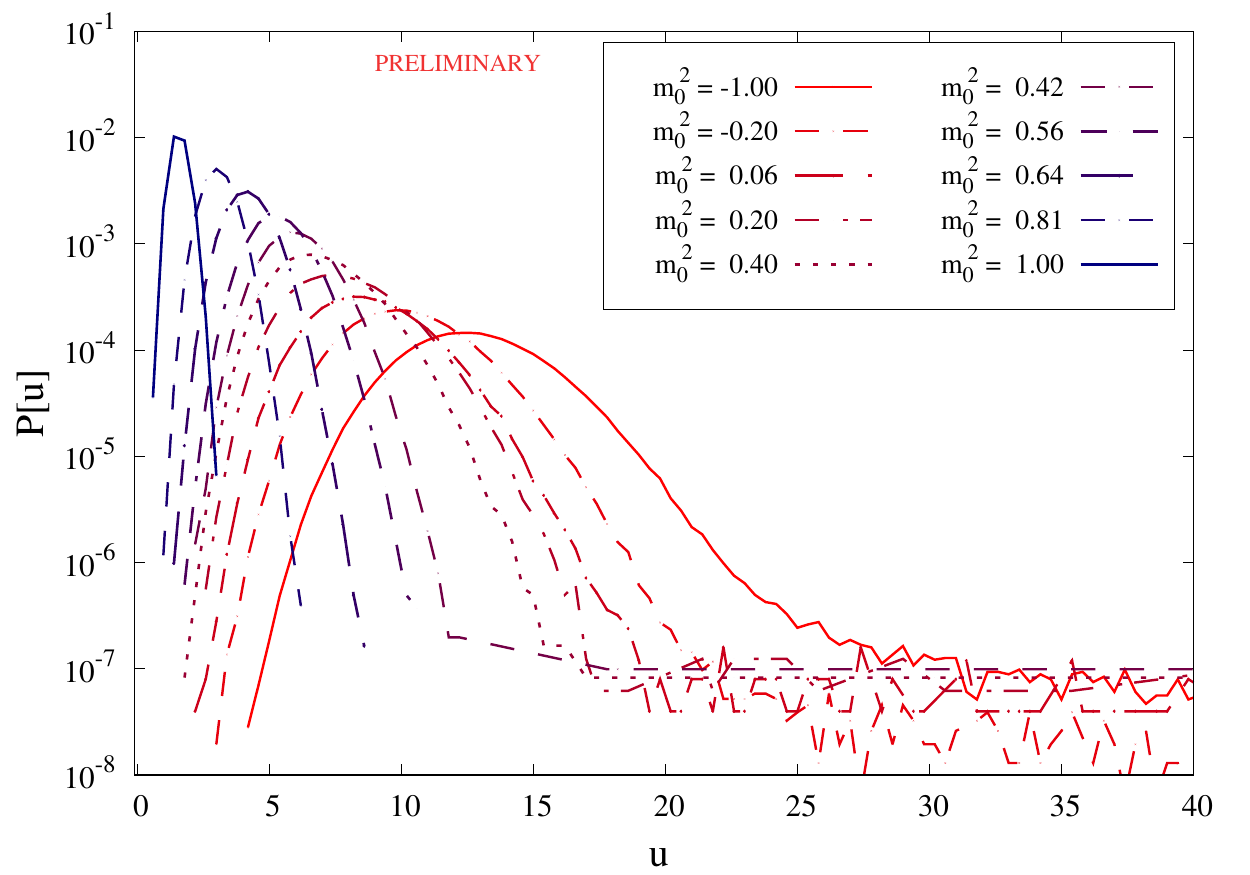}	
	
	\caption{$\mathcal{N} = 1$ Wess-Zumino model with a double-well superpotential. (Top) Langevin time histories of $\phi$ for various lattice mass $m_0^2$. The sign of the Pfaffian is also plotted for $m_0^2 = 0.16$. (Bottom-Left) Field $\langle \phi \rangle$, the sign of the Pfaffian $\langle {\rm sign} \left[ {\rm Pf} \mathcal{M}\right] \rangle$, and the Ward identity $\langle \mathcal{W} \rangle$ against lattice mass $m_0^2$. (Bottom-Right) Decay of the absolute drift for various lattice mass $m_0^2$. The plots are for a fixed lattice extent $L = 4$, and lattice coupling $\lambda_0 = 0.125$.}
	\label{fig:wz_quad}
\end{figure}

In Fig. \ref{fig:wz_quad} we show our simulation results. On the top panel, we have the scalar field $\phi$ (order parameter) against the Langevin time $\theta$ for various $m_0^2$ values, depicting the different phases of the theory. We have fixed the lattice extent to $L= 4$, and lattice coupling to $\lambda_0 = 0.125$. For $m_0^2 = +1$, the field configurations (blue squares) are confined, along with small fluctuations, to one of the classical vacua, $\phi = {m_0}/2{\lambda_0}$, implying that the theory is in an $\mathbb{Z}_2$ broken phase. At $m_0^2 = +0.16$, we observe the tunneling behavior, the field configurations (green triangles) undergo large fluctuations and they oscillate in between the two classical vacua at $\phi = \pm m_0/2\lambda_0$. The change in the sign of the Pfaffian (black diamonds) clearly illustrates this behavior. For $m_0^2 = -1$, the field configurations (red circles) suffer from small fluctuations around a single vacuum state, respecting the $\mathbb{Z}_2 $ symmetry. 

On the bottom left panel, we have the real part of the field $\langle \phi \rangle$ (order parameter), the sign of the Pfaffian $\langle {\rm sign}~{\rm Pf} \mathcal{M} \rangle$, and the simplest Ward identity $\langle \mathcal{W} \rangle$ against the lattice mass $m_0^2$ for a fixed lattice coupling $\lambda_0 = 0.125$. In the infinite-volume continuum theory, for large values of $m^2/\lambda$, the scalar field chooses a single unique ground state indicating a broken $\mathbb{Z}_2$ symmetry and unbroken SUSY in the model. We see that for mass larger than some critical value, $m_0^2 \ge m_{0, c}^2$, the scalar field (blue squares) selects the ground state $+ m_0/2\lambda_0$ and the sign of the Pfaffian (black diamonds) approaches $+1$. As $m_0^2$ is decreased, tunneling effects to the other vacuum state are observed, and the expectation value of the scalar field (green squares) vanishes $\langle \phi \rangle \sim 0$. This effect is a direct consequence of the Pfaffian flipping sign, reflected in $\langle {\rm sign}~{\rm Pf} \mathcal{M} \rangle \sim 0$. These results hint towards the restoration of $\mathbb{Z}_2$ symmetry and dynamical SUSY breaking. The above argument is supported by the Ward identity (yellow circles). As $m_0^2$ is decreased, we observe that the Ward identity no longer vanishes, that is, $\langle \mathcal{W} \rangle \neq 0$ indicating a transition from unbroken to broken SUSY phase. For $m_0^2 < 0$, we notice a $\mathbb{Z}_2$ symmetric phase with scalar field (red squares) $\langle \phi \rangle \sim 0$, and broken SUSY with $\langle \mathcal{W} \rangle \neq 0$. We show the decay of the absolute drift on the bottom right panel for our simulations. We observe exponential or faster decay for $m_0^2 > 0.42$ (illustrated by filled data points in the bottom left panel) and a power-law behavior for $m_0^2 \le 0.42$ (illustrated by unfilled data points in the bottom left panel). This could be pertaining to the singular drift problem, and we are looking further into it.

\subsection{Model with $PT$-symmetric  potential}

Our main goal is to cross-check the results obtained by Bender and Milton in Ref. \cite{Bender:1997ps}. There they have looked at a two-dimensional supersymmetric model with four supercharges with the superpotential $W(\phi) = - i \lambda (i \phi)^{(1+\delta)}$. Parity symmetry is broken in this supersymmetric model. The authors tried to answer the question on whether breaking of parity induces breaking of supersymmetry with the help of a perturbative expansion in parameter $\delta$. They found, through second order in $\delta$, that supersymmetry remained unbroken in the model, and suggested that SUSY could remain intact to all orders in powers of $\delta$. We plan to verify these results with the help of complex Langevin simulations of the model. We soon hope to report the results of ongoing simulations elsewhere \cite{paper}. 

\section{Conclusions}

In this work we have presented the preliminary results of our investigations on the two-dimensional scalar field theories with various interactions including the interesting cases of $PT$-invariant potentials. We laid out the lattice construction of the models and then studied the bosonic versions with $\phi^4$ and $PT$-symmetric potentials. After that we looked at a model with minimal supersymmetry, the two-dimensional $\cN = 1$ Wess-Zumino model. Our simulations for the model with double-well superpotential suggests that SUSY is preserved in this model when the mass parameter $m_0^2$ is greater than some critical value.

{\bf Acknowledgements:} We thank discussions with Takehiro Azuma and Navdeep Singh Dhindsa. The work of AJ was supported in part by the Start-up Research Grant (No. SRG/2019/002035) from the Science and Engineering Research Board (SERB), Government of India, and in part by a Seed Grant from the Indian Institute of Science Education and Research (IISER) Mohali. AK was partially supported by IISER Mohali and a CSIR Research Fellowship (Fellowship No. 517019).

\end{document}